\documentclass[nologo,11pt,a4paper]{ETHpaper}

\usepackage{tabularx}
\usepackage{MnSymbol}
\usepackage{booktabs}
\usepackage{caption}
\usepackage{hhline}
\usepackage[table,dvipsnames]{xcolor}
\usepackage{graphicx}
\usepackage[english]{babel}
\usepackage[numbers,sort&compress]{natbib} 
\usepackage{setspace}
\usepackage{subcaption}
\usepackage[capitalise]{cleveref}

\title{The nonlinear economy (I): \\ How resource constrains lead to business cycles}
\titlealternative{The nonlinear economy (I): How resource constrains lead to business cycles}

\author{Frank Schweitzer, Giona Casiraghi}
\authoralternative{F. Schweitzer, G. Casiraghi}
\address{Chair of Systems Design, ETH Zurich, Switzerland\\
  \url{www.sg.ethz.ch}}
\www{\url{http://www.sg.ethz.ch}}

\reference{(submitted for publication)
  } 
\makeframing

\usepackage{appendix}
\usepackage{amsmath} \usepackage[utf8]{inputenc}
\usepackage{xcolor} \usepackage{soul}
\usepackage{multirow}
\usepackage{hyperref}
\usepackage{xparse,tikz,calc} \usepackage{pifont}

\renewcommand{\epsilon}{\varepsilon}

\newcommand{\mean}[1]{\left\langle #1 \right\rangle}

\begin{document}

\maketitle

\begin{abstract}
  We explore the nonlinear dynamics of a macroeconomic model with resource constraints.
  The dynamics is derived from a production function that considers capital and a generalized  form of energy as inputs.
  Energy, the new variable, is depleted during the production process and has to be renewed, whereas capital grows with production and decreases from depreciation.
  Dependent on time scales and energy related control parameters, we obtain  steady states of high or low production, but also sustained oscillations that show properties of business cycles.
  We also find conditions for the coexistence of stable fixed points and limit cycles.
  Our model allows to specify investment and saving functions for Kaldor's model of business cycles.
  We provide evidence for an endogenous origin of business cycles if depleting resources are taken into account.
\end{abstract}

\hfill\emph{Dedicated to the memory of Jason Gallas}

\section{Introduction}
\label{sec:introduction}

For Jason Gallas nonlinear dynamics was not just a research domain, it was a lifelong passion. 
His impressive list of publications covers various application areas, ranging from geophysics to cancer.
Economics, however, was only touched indirectly, when studying bifurcations in competition models \citep{Gallas-2020-zip}. 
This is understandable.
Compared to the complex dynamics of, e.g., a Belousov-Zhabotinsky Reaction, one of Jason's favorite topics \citep{Gallas-2021-chiralitybelousov},  
nonlinear economic models are rather simple, and studying macroeconomic models with the arsenal of  nonlinear dynamics was fashionable during the 1980's to 2000 \citep{Semmler-1986-nonlineartheories,Lorenz-1989-nonlineardynamical,Puu-1993-nonlineareconomic}, not so much today.

We take the opportunity to change this perspective a bit with the hope to renew the attention of physicists and applied mathematicians for these kind of models.
Our aim is not to present a completely new approach.
After all, economists would not (yet?) see a need for this.
In this paper,
we first remind on how these macroeconomic models have been established, to then propose some extensions, and conclude by linking our approach to current research on active matter  in physics. 

A focus of our investigation is the resource dependence of production.
The issue itself is broadly discussed, for instance as ``resource dependence theory'' in management science \citep{Davis-Adam-2010-chapterresource}, but not as a model that could be  formally explored.
In economics, exhaustible resources play a role since Ricardo's time \citep{kurz2014ricardo,Erreygers-2009-hotellingrawlssolow}.
Notably, Hotelling made important contributions to formalize the discussion \citep{Hotelling-1931-economicsexhaustibleresources,devarajan1981hotelling,brazee2006reconciling}.
The main approach, however, is different from ours in that it primarily deals with how the timing of resource extraction affects its value and availability over time, balancing the diminishing stock with factors such as commodity prices, wages, profit rates, and demand.

Our starting point is the neoclassical growth model where production depends on the input of capital and labor.
But these are not modeled as resources that deplete during production.
Instead, they continuously grow: Labor force because of population growth, capital stock because of investments.
Therefore, we propose to consider a resource that is consumed during production.
We use the term ``energy'' for it, but interpret it very broadly as a natural resource.  
As a consequence, output is constrained by the availability and the renewal of this resource.
Growing production implies \emph{decreasing energy}.
This denotes an important difference to capital stock which is assumed to \emph{increase} with growing production.

Our second contribution is the formal derivation of a production \emph{dynamics}, starting from our production function.
In economics, this dynamics is often studied in so-called ``multiplier-accelerator'' models \citep{Mourao-Popescu-2022-revisitingmacroeconomic, Puu-1986-multiplier}.
  The multiplier describes the impact of an input, e.g., capital, on the expansion of production.
The accelerator describes the feedback of the growing output on the input variable, e.g., the growth of capital stock through the investment of a fraction of the output.
The dynamics assumes lags in this feedback process and can produce different types of steady-state solutions, including fix points, limit cycles, damped oscillations, but also unstable solutions, i.e., growing oscillations or even chaos. 

Theoretical economists, physicists and applied mathematicians have particularly studied a variant of such time-delayed dynamics, the Kalecki-Kaldor model \citep{Mehlab-Orlando-2023-bautinbifurcation,Matsumoto-Szidarovszky-2016-delaykaldor,Mircea-Neamtu-ea-2010-kaldorkalecki,Szydłowski-Krawiec-2001-kaldorkalecki,Szydłowski-Krawiec-2005-kaldorkalecki,Krawiec-Szydlowski-1999}. 
One of the reasons for this interest is the complex dynamics of the model.
Such a complexity is seen not as a drawback, but as an advantage, because it offers ample possibilities to generate a wealth of dynamic patterns.
These are considered a precondition to explain a complex real world phenomenon: \emph{business cycles} \citep{Mirowski-2015-birthbusiness,Semmler-1994-businesscyclesc,Gabisch-Lorenz-1987-businesscyclec}, originally dubbed \emph{trade cycles}. 
The question whether such delays are really needed to generate the dynamics of business cycles was  negatively answered already by \emph{Baron Kaldor} himself, who wrote in 1940: ``Previous attempts at constructing models of the Trade Cycle
- such as Mr. Kalecki's or Professor Tinbergen's - have thus
mostly been based on the assumption of statically stable situations,
where equilibrium would persist if once reached; the existence of
the cycle was explained as a result of the operation of certain
time-lags which prevented the new equilibrium from being reached,
once the old equilibrium, for some external cause, had been disturbed. In this sense all these theories may be regarded as
being derived from the `cobweb theorem'. The drawback of
such explanations is that the existence of an undamped cycle can
be shown only as a result of a happy coincidence, of a particular
constellation of the various time-lags and parameters assumed.
The introduction of the assumption of unstable positions of
equilibrium at and around the replacement level provides, however, [...] an explanation for a cycle of constant
amplitude irrespective of the particular values of the time-lags
and parameters involved. The time-lags are only important
here in determining the period of the cycle, they have no significance in explaining its existence.
Moreover, with the theories of the Tinbergen-Kalecki type,
the amplitude of the cycle depends on the size of the initial shock.
Here the amplitude is determined by endogeneous factors and the
assumption of 'initial shocks' is itself unnecessary.''\citep{Kaldor-1940-modeltrade}

This long quotation sets a nice stage for our own investigations.
In line with the cited research, we attempt to explain business cycles as \emph{endogenously} created by coupled nonlinear dynamics.
This contrasts other explanations of business cycles as the result of \emph{exogenous} perturbations of an otherwise stable dynamics. 
But differently from the cited research, we will not utilize delayed differential equations to generate cycles, nor propose \emph{ad hoc} nonlinear functions.
Instead, we will derive the non-linear dynamics from the production function, using suitable assumptions for the dynamics of capital and energy.
This will shed new light on the formal preconditions for obtaining limit cycle dynamics.

\section{A macroeconomic growth model}
\label{sec:macro-growth}

\subsection{Production function}
\label{sec:production-function-1}

Economic models often start from a 
production function
\begin{align}
  \label{eq:22}
\hat{Y}[X_{1}(t),X_{2}(t),....] = Y_{0}+ {Y}=Y_{0}+ \mathcal{F}[X_{1}(t),X_{2}(t),....]
\end{align}
$\hat Y$ denotes the output, or production, of a macroeconomic entity.
For instance, the GDP (gross domestic product) would be an output measure for a country.
It is important to note that $\hat Y$ is the \emph{output per time unit} $\Delta t$, e.g., one quarter or one year.
So, physically speaking it is a \emph{velocity}.
Consequently, the time derivative $dY/dt$ discussed below is analogous to an \emph{acceleration} and the r.h.s. of the dynamics specifies economic \emph{forces}. 

The variables $X_{1}(t),X_{2}(t),....$ denote different inputs, e.g., capital, labor, resources, and the function $\mathcal{F}[\cdot]$ describes how the input is transformed into a valuable output, similar to the alchemistic idea of transforming lead into gold.
The normalization $Y_{0}$ can be seen as an equilibrium state with baseline economic activity.  

Putting \cref{eq:22} to use requires to specify (i) the input variables $X_{i}$, (ii) their combination in a \emph{nonlinear} function $\mathcal{F}[\cdot]$ and (iii) their possible dynamics, $X_{i}(t)$.
Let us first solve issue (ii).
An additive combination of input variables $\mathcal{F}=a_{1}X_{1}+a_{2}X_{2}+...$ implies that inputs can be \emph{substituted} to some degree, i.e., a shortage of $X_{1}$ can be compensated by an increase of $X_{2}$.
A multiplicative combination,  $\mathcal{F}=X^{a_{1}}_{1}\cdot X_{2}^{a_{2}}\cdots$, on the other hand highlights that inputs are essential:  $X_{1}$ cannot be completely substituted by $X_{2}$. 
The exponents $a_{1}$, $a_{2}$ denote \emph{elasticities}, i.e.,  \emph{relative changes} of output in response to relative changes of input:
\begin{align}
  \label{eq:24}
  a_{i}=\frac{\partial Y/Y}{\partial X_{i}/X_{i}}=\frac{X_{i}}{Y}\frac{\partial Y}{\partial X_{i}}=\frac{\partial \ln Y}{\partial \ln X_{i}}
\end{align}

The question is how much freedom one has in choosing the nonlinear function.
As we will see, the functional form is quite restricted by some fundamental assumptions.
For a general derivation, we refer to \citep{Arrow-Chenery-ea-1961-capitallaborsubstitutiona}, whereas we follow the more didactic approach of \citep{Koch-2013-cobbdouglasfunction}.
Let us  consider only two inputs $X_{1}$, $X_{2}$. Hence, we need to determine $Y(X_{1},X_{2})$.
The first fundamental assumption is to consider only \emph{homogeneous} functions with the power $n$ known as the degree of homogeneity: 
\begin{align}
  \label{eq:31}
  Y(\alpha X_{1},\alpha X_{2})=\alpha^{n} Y(X_{1},X_{2})
\end{align}
This relation plays a role when discussing so-called \emph{returns to scale} in economics.
If we would increase the two inputs by an arbitrary factor $\alpha$, then the output increases by a factor $\alpha^{n}$.
Let us consider linear homogeneity, $n=1$.
Then the production function can be expressed in terms of partial derivatives by means of the Euler theorem:
        \begin{align}\label{eq:32}
          1\cdot Y(X_{1},X_{2})= X_{1}\frac{\partial Y(X_{1},X_{2})}{\partial X_{1}}+ X_{2}\frac{\partial Y(X_{1},X_{2})}{\partial X_{2}}
        \end{align}
The second fundamental assumption is about \emph{diminishing} returns to scale.
It means that the output still increases with increasing input. However, its impact becomes smaller and smaller.
This reflects economic reality: it does not make sense to scale up production beyond certain limits because the marginal product tends to zero.
Formally: 
      \begin{align}\label{eq:33}
        \frac{\partial^{2} Y(X_{1},X_{2})}{\partial X_{1}^{2}}<0\;;\;
        \frac{\partial^{2} Y(X_{1},X_{2})}{\partial X_{2}^{2}}<0
      \end{align}
The third fundamental assumption is about the independence of the inputs, which allows a \emph{separation of variables}: $Y(X_{1},X_{2})=G(X_{1})H(X_{2})$
      \begin{align}\label{eq:34}
        G(X_{1})H(X_{2})&=X_{1}\frac{d G(X_{1})}{d X_{1}}H(X_{2})+ X_{2}\frac{d H(X_{2})}{d X_{2}}G(X_{1}) \nonumber \\
1&=X_{1}\frac{d G(X_{1})/d X_{1}}{G(X_{1})}+ X_{2}\frac{d H(X_{2})/d X_{2}}{H(X_{2})}
      \end{align}
\cref{eq:34}  can only hold if 
      \begin{align}\label{eq:35}
        X_{1}\frac{d G(X_{1})/d X_{1}}{G(X_{1})}&=a_{1}\;;\quad X_{2}\frac{d H(X_{2})/d X_{2}}{H(X_{2})}=a_{2} \\
        a_{1}+a_{2}&=1
      \end{align}
Integration then leads to 
      \begin{align}\label{eq:36}
        \int  \frac{d G(X_{1})/d X_{1}}{G(X_{1})} dX_{1} & =\int \frac{a_{1}}{X_{1}} dX_{1}
        = a_{1} \ln X_{1}+C_{1} \nonumber \\
        \int  \frac{d H(X_{2})/d X_{2}}{H(X_{2})} dX_{2} & =\int \frac{a_{2}}{X_{2}} dX_{2}
        = a_{2} \ln X_{2}+C_{2}
      \end{align}
With the initial condition  $Y(1,1)=A=e^{(C_{1}+C_{2})}$ one eventually finds as the functional form for the production function:
      \begin{align}\label{eq:37}
{Y(X_{1},X_{2})=A X_{1}^{a_{1}}X_{2}^{a_{2}}}
      \end{align}
We will use this form in the following. 
The pre-factor $A$ is known as the \emph{total factor productivity} and describes how efficient the inputs are used, e.g., by an advanced technology.
In general, $A$ accounts for effects in total output not caused by inputs, for instance the impact of good weather on agricultural output.
It should be noted that today increasing production is mostly attributed to improvements in the total factor productivity \citep{Hulten-2000-totalfactorproductivity}.

\subsection{Cobb-Douglas production function}
\label{sec:cobb-dougl-prod}

We now solve issue (i), i.e., we specify the  input variables.
To discuss a concrete example, we refer to the \emph{neoclassical growth model} \citep{Solow-2001-fromneoclassical,Solow-Tobin-ea-1966-neoclassicalgrowth,Solow-1956-contributiontheory}, a 
standard macroeconomic model that uses capital $K(t)$ and  labor  $L(t)$ as inputs.
These variables are combined in a so-called Cobb-Douglas production function \citep{Cobb-Douglas-1928-theoryproduction,Zellner-Kmenta-Dreze-1966-specificationestimation,Heathfield-1971-cobbdouglasfunction}:
\begin{align}
  \label{eq:23}
  \hat{Y}={Y}(K,L)=A\; K^{1-\alpha}L^{\alpha}
\end{align}
which neglects $Y_{0}$. 
The two exponents denote the elasticities with respect to labor and capital: 
\begin{align}
  \label{eq:27}
  \alpha=\frac{\partial \ln Y}{\partial \ln L}\;; \quad
\beta=1- \alpha=\frac{\partial \ln Y}{\partial \ln K}
\end{align}
For the dynamics of the input variables 
the \emph{neoclassical growth model} assumes: 
\begin{align}
  \label{eq:25}
  \frac{d K}{dt}=sY-\kappa K\;;\quad \frac{d L}{dt}=rL 
\end{align}
The capital stock $K$ grows via an investment $I=sY$ that is coupled to the current production,  $0<s<1$ being the savings rate.
I.e., by means of capital there is a positive feedback between the current and the future output level.
$\kappa$ is the depreciation rate, i.e., the value of capital stock exponentially decays if it is not maintained.
For the labor force $L$ an exponential growth is assumed, which is inspired by population dynamics.   
If the net growth rate $r>0$, then births and immigration dominate and result in an exponential increase, if $r<0$, then deaths and emigration dominate and result in an exponential decay of the population size which is equal to the available labor force.

For the output $Y[K,L]$, an instantaneous adjustment is assumed.
Instead of $dY/dt$, the economic model only considers $dK/dt$ and $dL/dt$ and postulates that $Y$ takes its new level immediately after $K$ and $L$ change.
In physics, this is known as a \emph{separation of time scales}.
Compared to the slow change of $K$ and $L$, $Y$ changes fast, therefore it can be assumed in \emph{quasi-stationary} equilibrium. 
This reduces the discussion to a comparison of the different values of $Y$ \emph{before} and \emph{after} changes of $K$ and $L$. 

This setup has become the canonical model for the  exogenous explanation of business cycles \citep{King-Plosser-ea-1988-productiongrowth}.
Subsequent works have introduced additional assumptions to endogenize the causes of business cycles \citep{Pangallo-2023-synchronizationendogenous,Kroujiline-Gusev-ea-2021,Farmer-2014-theevolution,Gallegati-Gardini-ea-2003-hickstradea}.

\subsection{Comparative statics}
\label{sec:comparative-statics}

The simplifying assumptions of the neoclassical growth model consequently generate only a 
very simple dynamics, namely either exponential growth or exponential decay of output.
Therefore, the modeling aim is not to study the dynamics, but only the stationary state obtained for the \emph{reduced variable} $y=Y/L$: \emph{output per capita}.
For instance, GDP per capita is an important economic indicator to compare the wealth of countries.
Dividing \cref{eq:23} and \cref{eq:25} by $L$ results, after some straightforward transformations, into the set of equations for the reduced variables 
\begin{align}
  \label{eq:28}
  y=\frac{Y}{L}=A k^{\beta}\;;\quad \dot{k}=\frac{d}{dt}\left(\frac{K}{L}\right)=sy-(r+\kappa)k
\end{align}
These equations make it obvious that large immigration rates $r$ act similar to large depreciation rates $\kappa$, drastically reducing the output per capita, $y$, and, hence, the capital per capita, $k$, available in a given country. 
More important from a modeling perspective, the dynamics now always reaches a stable equilibrium state, $k_{\star}$, for the capital per capita. 
From $\dot{k}=0$ we obtain:
\begin{align}
  \label{eq:29}
  s A k_{\star}^{\beta}=(r+\kappa)k\;;\quad k_{\star}=\left[\frac{sA}{r+\kappa}\right]^{\frac{1}{\alpha}}
\end{align}
A government should focus its policy design on the optimal value $s_{\mathrm{gold}}$ of the savings rate, i.e., the optimal split of the output in investment and consumption.  
$I=sY$ is the fraction of output reinvested into the growth of the capital stock and, hence, the further growth of the economy.
Therefore, only the remainder of the output $C=(1-s)Y$ is left for consumption, e.g., increasing pensions by the government.
Consumption per capita in the equilibrium state $k_{\star}$ is given as $c_{\star}=(1-s)y_{\star} = A k_{\star}^{\beta}-(r+\kappa)k_{\star}$ and maximizing consumption  means:  
\begin{align}
  \label{eq:30}
  \frac{dc_{\star}}{dk_{\star}}&=0 \;; \quad s_{\mathrm{gold}}=(1-\alpha)  \;; \quad k_{\mathrm{gold}}=
                                          \left[\frac{(1-\alpha)A}{r+\kappa}\right]^{\frac{1}{\alpha}}
\end{align}
The policy recommendation for governments, according to the neoclassical growth model, is then to increase the savings rate $s$ if it is below $s_{\mathrm{gold}}$ because this will increase both the national wealth, i.e., the capital per capita, \emph{and} the consumption.
If, however, $s$ is larger than $s_{\mathrm{gold}}$, then the recommendation is to \emph{decrease} the savings rate i.e., to lower the national wealth at the expense of increasing consumption.
This is not the place to discuss the validity of such policy implications.
But OECD recommendations for  wealthy countries to increase governmental expenditures  are fueled by such insights.

\section{Coupling between production and energy}
\label{sec:coupl-betw-prod}

\subsection{Modifications}
\label{sec:modifications}

``The greater the prestige, the greater the opposition'' also applies to the neoclassical growth model.
Instead of reviewing the many criticisms and the various economic debates that followed, we will concentrate on some nonlinear dynamic aspects.
To introduce these, we replace labor force as the relevant input variable.
There were certainly periods in history where economic growth was predominantly driven by an exponential increase of the \emph{population}.
But nowadays this growth does not easily translate into exponential growth of \emph{labor force}.
A refined dynamics of $L(t)$, \cref{eq:25} should also reflect labor related issues such as unemployment, unskilled labor and working poor, lack of specialized workforce, etc. which are not further discussed here \cite{Farmer-2014-theevolution}. 

Instead, we consider, in addition to capital, $K(t)$, a different input variable, \emph{energy}, $E(t)$.
It is a general form of ``energy'' to reflect also other material resources needed for production.
Considering resources that are \emph{depleted} denotes a conceptual change.
The production function $Y[K,L]$ uses capital and labor as essential inputs, but the process of production does \emph{not reduce} any of the inputs.
Thus, $K$ and $L$ act as \emph{catalysts} for the production, very similar to catalysts in chemical reactions. 
They are needed for the ``reaction'', but are not consumed during the production.
The input variable $E$ however is consumed, i.e., the initial resource is diminished.

To better understand the consequences of this modification, we remind that production $Y$ is a \emph{velocity}, output per time unit.
However, capital $K$ and energy $E$ are not \emph{flow variables} in the system dynamic sense, i.e., quantities per \emph{time unit}, but \emph{stock variables} that can be accumulated or depleted.  
Negative values of capital would indicate debt, which is possible in principle but will not be considered here.
To avoid negative values, we constrain these stock variables by  \emph{floor values} $K_{f}\geq 0$, $E_{f}\geq 0$ such that the input variables remain positive in the dynamic case. 
Hence, our production function reads $Y[K(t),E(t)]$ and has the general form already discussed  in \cref{sec:production-function-1}:
\begin{align}
  \label{eq:49}
  \hat{Y}[K(t),E(t)]=Y_{0}+{Y}=Y_{0}+ A\, \left[K(t)-K_{f}\right]^{a_{K}} \left[E(t)-E_{f}\right]^{a_{E}}\;;\; a_{K}+a_{E}=1
\end{align}
Here, we have considered a baseline output $Y_{0}>0$ for the case that no additional capital or energy is used.
This value shall reflect basic economic activities, which are always present, so $Y_{0}$ is not zero as in the neoclassical growth model.
The aim of all economies is to reach a level of production well above $Y_{0}$, by means of capital and energy.
Hence, our production function contains of two terms where the second one reflects changes in production resulting from the input of $K$ and $E$. 
The fact that the input variable $E$ is consumed is the precondition for \emph{increasing} production, i.e., decreasing energy has a positive effect.

\subsection{Eigendynamics and driven dynamics}
\label{sec:eigendyn-driv-dynam}

To contribute to economic modeling we introduce an explicit dynamics for production.
This point deserves a broader discussion.
Our reference point, the neoclassical growth model, assumes only a dynamics for the input variables, to recalculate the output instantaneously at every time step.
So $Y$ simply follows the dynamics of $K$ and $L$.
We instead consider an explicit dynamics ${d \hat Y}/{dt}$ resulting from two terms, the \emph{eigendynamics} $P[Y]$ and the \emph{driven dynamics} $Q[K,E]$:
\begin{align}
  \label{eq:63}
  \frac{d \hat Y}{dt}= \frac{d Y[Y,K,E]}{dt} = P[Y]+Q[K,E]
\end{align}
Eigendynamics refers to changes in the production that only depend on $Y$, but not on input of capital and energy.
We already proposed that in such a case only the baseline production $Y_{0}$ should be reached.
Hence, our eigendynamics $P[Y]$ has to reflect how this baseline value is established.
This can be realized by different forms of saturation dynamics.
To be flexible, we choose a very general ansatz: 
\begin{align}
  \label{eq:62}
   P[Y] = g_{1}\left[Y_{0}-Y(t)\right]+g_{2} Y(t)\left[Y_{0}-Y(t)\right] 
\end{align}
This is known as the mixed source model in management science.
The first term solves the so-called ``cold start problem'', i.e., it guarantees that production can be induced initially, without pre-existing production.
The second term reflects the fact that existing production has a positive, i.e., amplifying, impact on the further growth of output.
Both terms saturate at a level $Y_{0}$, and the parameters $g_{1}$, $g_{2}$ determine the specific shape of the growth dynamics.
They only affect how fast the baseline production is reached, but do not determine $Y_{0}$.

The driven dynamics $Q[K,E]$ reflects changes of production resulting from the dynamic input of capital and energy.
For the derivation we can 
use Euler's Theorem, starting from \cref{eq:32} with $\tilde K= K(t)-K_{f}$, $\tilde E= E(t)-E_{f}$ as our input variables $X_{j}(t)$:
\begin{align}
  \label{eq:300}
Q[K,E]&= \sum_{j} \frac{dY_{j}}{dt} = \sum_{j} \frac{\partial Y}{\partial X_{j}} \frac{1}{\epsilon_{j}} \frac{dX_{j}}{dt} =
        \frac{\partial Y}{\partial \tilde K} \frac{1}{\epsilon_{K}} \frac{d\tilde K}{dt}+
        \frac{\partial Y}{\partial \tilde E} \frac{1}{\epsilon_{E}} \frac{d\tilde E}{dt} 
\end{align}
In \cref{eq:300} we have introduced two different time scales, $dt_{K}=\epsilon_{K}dt$ and $dt_{E}=\epsilon_{E}dt$ to allow the dynamics of $Y(t)$ to evolve on a time scale different from the dynamics of $K(t)$ and $E(t)$.
$dt\equiv dt_{Y}$ then refers to the dynamics of $Y(t)$.  
The relevance of these different scales will be demonstrated below.

To complete the formal description, we need to determine $\partial Y/\partial \tilde K$ and $\partial Y/\partial \tilde E$ and provide kinetic assumptions for $d \tilde K/dt$ and $d\tilde E/dt$.
In line with the use of energy as a depleting resource the dynamics of $K$ and $E$ can only be a \emph{dissipative} dynamics, i.e., it has the general form:
\begin{align}
  \label{eq:56}
\frac{dX_{i}(t)}{dt}&=-\gamma_{i} \left[X_{i}-X_{f}\right] + Q_{i}   
\end{align}
The damping term $-\gamma_{i} [X_{i}-X_{f}]$ ensures, on the one hand, that an unbounded growth is prevented and, on the other hand, that the value of the input variable stays above a \emph{floor value} $X_{f}\geq 0$, i.e., remains positive.
Then, the dynamic variable $X_{i}(t)$ converges to a stationary value,  $[X_{i}^{\star}-X_{f}]=Q_{i}/\gamma_{i}$.
The source term $Q_{i}$ has to compensate dissipation, i.e., it denotes the growth or the inflow of resources.
Hence, instead of a classical \emph{conservative} system, we model a \emph{dissipative} system. 
From a physical perspective the economy  is a pumped system, similar to active matter, a point we will further elaborate in the discussion.

\subsection{Capital input}
\label{sec:capital-input}

To specify the dynamics of the input variables   
we start with  capital as the ``classic'' input variable.
Using the general form of the production function, \cref{eq:49}, with $\tilde K=K(t)-K_{f}$, $\tilde E=E(t)-E_{f}$ and $a_{K}=1/2$,  gives 
\begin{align}
  \label{eq:90}
  \frac{\partial Y}{\partial \tilde K}=\frac{A}{2}\tilde K^{-1/2}\tilde E^{1/2}= \frac{Y}{2\tilde K}\end{align}
For the dynamics of capital we use the general dissipative ansatz, \cref{eq:56}, but need  to consider that it evolves at the time scale $t_{K}=\epsilon_{K}t$:
\begin{align}
  \label{eq:5a}
\frac{1}{\epsilon_{K}}  \frac{d \tilde K}{dt}=-\kappa \left[K(t)-K_{f}\right] + Q_{K}
  \;;\quad Q_{K}=s{Y}(t) \;;\quad {\tilde K}=\frac{s}{\kappa}Y  
\end{align}
For the source term $Q_{K}$ we have re-used the assumption from the neoclassical growth model, \cref{eq:25}, i.e., a share $s$ of the total output is invested into capital stock.
The stationary solution of \cref{eq:5a}  gives us a \emph{linear} relation between $\tilde K$ and $Y$. 
For the dynamics of production with reference to $K$ we then obtain:
\begin{align}
  \label{eq:160}
\frac{dY_{K}}{dt}= \frac{\partial Y}{\partial \tilde K}\frac{1}{\epsilon_{K}} \frac{d \tilde K}{dt}&= \left(\frac{\kappa}{2s}\right) \frac{1}{\epsilon_{K}} \frac{d\tilde K}{dt}
                                                 = \frac{\kappa}{2} Y(t)-\frac{\kappa^{2}}{2s} \left[K(t)-K_{f}\right]
\end{align}

\subsection{Energy input}
\label{sec:energy-input}

For the second input variable, energy, we obtain from the production function, \cref{eq:49}, with $\tilde K=K(t)-K_{f}$, $\tilde E=E(t)-E_{f}$ and $a_{E}=1/2$
\begin{align}
  \label{eq:9}
  \frac{\partial Y}{\partial \tilde E}=\frac{A}{2}\tilde K^{1/2}\tilde E^{-1/2}=\frac{Y}{2\tilde E} \end{align}
To specify the dynamics for $\tilde E(t)$ at the time scale $t_{E}=\epsilon_{E}t$, we
start from a power series ansatz \citep{Schweitzer-2019}: 
\begin{align}
  \label{eq:64}
  \frac{d \tilde E}{dt_{E}}=Q_{E}+\tilde E \sum_{n=0}^{m}d_{n}Y^{n}=Q_{E}+d_{0}\tilde E+d_{1}\tilde E Y+d_{2}\tilde E Y^{2}... 
\end{align}
which generalizes the dissipative dynamics, \cref{eq:56}.
For $Q_{E}$ we consider that energy 
is provided at a constant rate $q$, like sun radiation or a steady supply of fossil fuels.
The term without $Y$ has to reflect the dissipation, i.e., $d_{0}=-c$, where $c$ is the dissipation rate.
If we restrict the power series to $n=2$, the two remaining terms just describe a saturation dynamics if $d_{2}=-\zeta$ is negative:
\begin{align}
  \label{eq:65}
  d_{1}\tilde E Y-\zeta \tilde E Y^{2}=\zeta \tilde E \ Y\left[Y_{s}-Y\right] \;;\quad Y_{s}=\frac{d_{1}}{\zeta}
\end{align}
This is the same saturation dynamics as assumed for second term in the mixed source model.
It results: 
\begin{align}
  \label{eq:555}
  \frac{1}{\epsilon_{E}} \frac{d \tilde E}{dt}=q-c\tilde E + \zeta \tilde E \left[Y_{s}Y-Y^{2}\right]
  \end{align}
It makes sense to combine the two negative terms in \cref{eq:555} in a generalized dissipation term $-\gamma_{E}E$ with
\begin{align}
  \label{eq:60}
  \gamma_{E}=c+\zeta Y(t)^{2}
\end{align}
In addition to the exponential decay of energy at a rate $c$, the dissipation function $\gamma_{E}$ reflects that
\emph{any} change of production, \emph{positive or negative}, requires to consume additional energy \citep{Erdmann-Ebeling-ea-2000-brownianparticles,Ebeling2003}.
Therefore, it is not a constant, but depends on the \emph{squared} change of production, $Y(t)^{2}$.
The parameter $\zeta$ captures the inefficiency in using the resource $E$ to boost production.
The higher $\zeta$, the more the energy is reduced to change production by a given amount.

For $Y(t)\to 0$, i.e., no production, we obtain from \cref{eq:555} in the stationary limit the source value for energy,
$E_{Q}-E_{f}=q/c$. 
For $Y(t)\to Y_{0}$, instead, we regain our baseline production $Y_{0}[K_{0},E_{0}]$ with:
\begin{align}
  \label{eq:11}
  \left[E_{0}-E_{f}\right]=\frac{q}{c-\zeta Y_{0}[Y_{s}-Y_{0}]}
\end{align}
The dynamics of production with reference to $E$ eventually reads as:
\begin{align}
  \label{eq:10}
  \frac{dY_{E}}{dt}= \frac{\partial Y}{\partial \tilde E} \frac{1}{\epsilon_{E}}\frac{d\tilde E}{dt}&=\frac{Y}{2[E(t)-E_{f}]}{q}- Y\frac{c}{2}+Y^{2}\frac{\zeta Y_{s}}{2}- Y^{3}\frac{\zeta}{2}
\end{align}

\subsection{Oscillations vs fixed points}
\label{sec:oscill-vs-fixed}

Before introducing additional assumptions, we investigate the full dynamics.
In particular, we have to consider the eigendynamics $P[Y]$,  \cref{eq:63},  of $\hat Y$ and the contributions $Q[K,E]$, \cref{eq:300} from the input variables.
Putting our separate equations together at time scale $t$, we arrive at the three coupled nonlinear equations:
\begin{align}
  \label{eq:17}
   \frac{d\hat Y}{dt} =   & = Y(t)\left[\frac{\kappa-c}{2} +\frac{q}{2[E(t)-E_{f}]}-g_{1}+g_{2}Y_{0}  \right]  
+Y(t)^{2}\left[\frac{\zeta Y_{s}}{2}-g_{2}\right] -Y(t)^{3}\frac{\zeta}{2} \nonumber \\
& \quad
  - \frac{\kappa^{2}}{2s} \left[K(t)-K_{f}\right]+g_{1}Y_{0}                          \\
  \label{eq:661}
  \frac{d \tilde E}{dt}=\frac{dE(t)}{dt} &= \epsilon_{E} q-\epsilon_{E} \left[E(t)-E_{f}\right]
                                           \left[c+{\zeta Y_{s}} Y(t)-{\zeta} Y(t)^{2}\right] \\
  \label{eq:662}
\frac{d\tilde K}{dt}=\frac{dK(t)}{dt} &= \epsilon_{K} s  Y(t) - \epsilon_{K} \kappa \left[K(t)-K_{f}\right] \end{align}
\begin{figure}[htbp]
  \centering
  \includegraphics[width=0.29\textwidth]{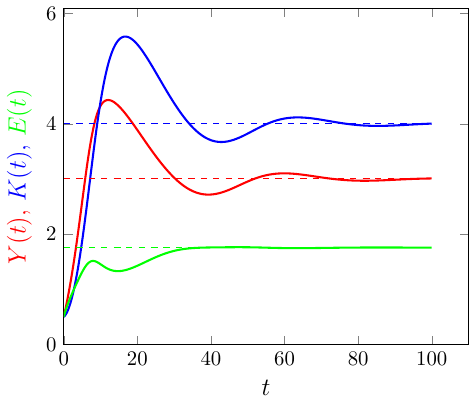}(a) \hfill
  \includegraphics[width=0.29\textwidth]{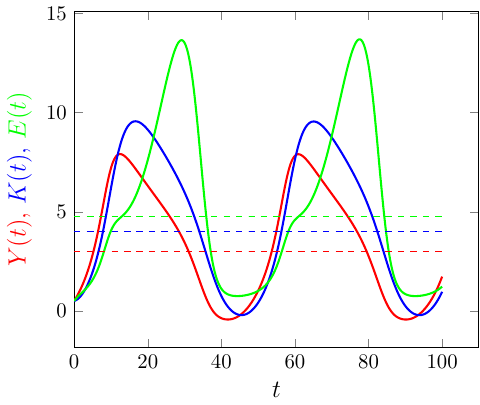}(b) \hfill
  \includegraphics[width=0.29\textwidth]{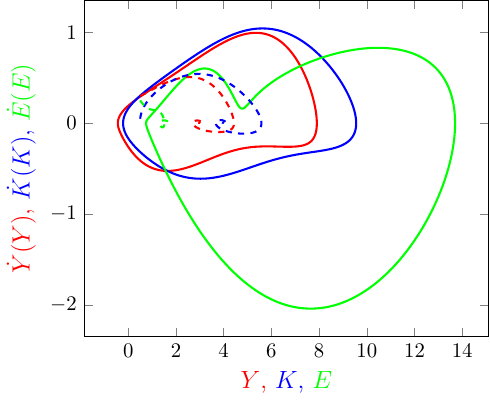}(c) \hfill 
  \caption{Production $Y(t)$ (green), capital $K(t)$ (blue) and energy $E(t)$ (green) for (a) $\zeta$=0.04 and (b) $\zeta$=0.02. The dashed lines give the respective baseline values $Y_{0}$=3, $K_{0}$, \cref{eq:5a}, $E_{0}$, \cref{eq:11}. 
    (c) Phase plots $\dot{Y}(Y)$ (red),  $\dot{K}(K)$ (blue) , $\dot{E}(E)$ (green) for $\zeta$=0.02 (solid) and $\zeta$=0.04 (dashed). Other parameters: $d_{1}$=0.225, $s$=0.8, $q$=0.5, $c$=0.6, $\kappa$=0.6, $Y_{0}$=3, $K_{f}$=0, $E_{f}$=0, $g_{1}$=0.05, $g_{2}$=0.01, $\epsilon_{K}$=0.5, $\epsilon_{E}$=1.}
  \label{fig:yk-kt-0100-01679}
\end{figure}

We solve this dynamics numerically.
As illustrated in \cref{fig:yk-kt-0100-01679} we find two significantly different outcomes; (i) a stationary production and (ii) sustained oscillations.
Unfortunately, the stationary solution is $Y(t)\to Y_{0}$, i.e., after some intermediate oscillations we are back at the baseline scenario, while looking for a case with $Y(t)\gg Y_{0}$.
This is obtained in the second scenario where we can verify that  $\mean{Y(t)}>Y_{0}$.
Thus, the average production is indeed above the baseline.
More important, during certain time periods $Y(t)$ is much larger that $Y_{0}$, i.e., we see a \emph{boom} phase of the economy. 
But this does not last long, and is followed by a steep decline of production that can even reach negative values, very similar to \emph{business cycles}.
These comprise \emph{four phases} of \emph{different duration}, (a) a \emph{short} boom phase, (b) a \emph{long} recession phase, (c) a \emph{short} depression phase, and (d) a \emph{long} recovery phase, after which a new cycle starts.
This is indeed captured with our dynamics of production.

\section{Nonlinear oscillations and business cycles}
\label{sec:nonl-oscill-busin}

\subsection{Two-dimensional dynamics}
\label{sec:two-dimens-dynam}

In order to calculate a bifurcation diagram it would be convenient to reduce the full dynamics of three coupled variables to two variables.
The simplest way of doing so is to assume that the dynamics of one of the input variables, $K$ or $E$, relaxes very fast and therefore can be described by it's \emph{quasistationary} equilibrium.
This is equivalent of choosing $\epsilon_{K}\to 0$ in \cref{eq:662} or $\epsilon_{E}\to 0$ in \cref{eq:661}, which results in $dK/dt=0$ or $dE/dt=0$.
Note that this does not imply $K-K_{f}=0$ or $E-E_{f}=0$, instead the respective variables are tidily coupled to the production $Y$ and therefore can still change over time via $Y(t)$.
We obtain as quasi-stationary equilibria, $K_{\star}$, $E_{\star}$:
\begin{align}
  \label{eq:19}
  K_{\star}-K_{f}=\frac{s}{\kappa}Y(t)\;; \quad 
  E_{\star}-E_{f}=\frac{q}{c-\zeta Y(t)\left[Y_{s}-Y(t)\right]}
\end{align}
If we choose the quasi-stationary approximation for capital, $K(t)=K_{\star}$, this gives the following two coupled equations for $Y(t)$ and $E(t)$: 
\begin{align}
  \label{eq:17a}
   \frac{d\hat Y}{dt}    & = Y(t)\left[-\frac{c}{2} +\frac{q}{2[E(t)-E_{f}]}-g_{1}+g_{2}Y_{0}  \right]  
                            +Y(t)^{2}\left[\frac{\zeta Y_{s}}{2}-g_{2}\right] -Y(t)^{3}\frac{\zeta}{2}
  +g_{1}Y_{0}                          \\
  \frac{d \tilde E}{dt} &= \epsilon_{E} q-\epsilon_{E} \left[E(t)-E_{f}\right]
                                           \left[c+{\zeta Y_{s}} Y(t)-{\zeta} Y(t)^{2}\right] 
\end{align}
This reduced dynamics still gives us the same characteristic patterns as before if we use, e.g., $\epsilon_{E}$ as control parameter: (a) for larger $\epsilon_{E}$ damped oscillations of production that converge to $Y_{0}$ over time, (b) for smaller $\epsilon_{E}$ sustained oscillations that persist over time. 

If we choose instead the quasi-stationary approximation for energy, $E(t)=E_{\star}$, we have 
that production only depends on capital
\begin{align}
  \label{eq:21}
  \frac{d\hat Y[K]}{dt} & = g_{1}\left[Y_{0}-Y(t)\right] +g_{2}Y(t)\left[Y_{0}-Y(t)\right]+ \frac{\kappa}{2}Y(t)                       - \frac{\kappa^{2}}{2s} \left[K(t)-K_{f}\right] \\
  \frac{d\tilde K}{dt} &= \epsilon_{K} s  Y(t) - \epsilon_{K} \kappa \left[K(t)-K_{f}\right] \nonumber
\end{align}
In this case, we only obtain damped oscillations of production that converge to $Y_{0}$ over time, which is not surprising because we lack the nonlinear feedback from the energy resource. 

To avoid this trivial scenario we develop a different limit case of eliminating $E$.
We verify in \cref{eq:17} that the coupling between the dynamics of production and of energy is solely given by the term $Y q/(2\tilde E)$.
Using the production function, we have an expression for the energy as a function of capital: 
\begin{align}
  \label{eq:51}
\frac{\partial Y}{\partial \tilde K}= \frac{A}{2}\tilde K^{-1/2}\tilde E^{1/2}=\frac{\kappa}{2s} \;;\quad \tilde E=\frac{\kappa^{2}}{A^{2}s^{2}}\tilde K
\end{align}
Together with $Y/\tilde K=\kappa/s$ 
this can be used to replace
\begin{align}
  \label{eq:58}
\frac{Y}{2\tilde E}{q}=\frac{A^{2}s^{2}}{2\kappa^{2}}\frac{Y}{K}{q}=\frac{s A^{2}q}{2\kappa} 
  \end{align}
We note that the parameter $A^{2}$ is related to $Y^{2}_{0}=A^{2}\tilde K_{0}\tilde E_{0}$ with $\tilde K_{0}$ given in \cref{eq:5a} and $\tilde E_{0}$ given in \cref{eq:11}.
  Hence,
  \begin{align}
    \label{eq:59}
    A^{2}=\frac{Y_{0}\kappa}{s q}(c-\zeta Y_{0}[Y_{s}-Y_{0}])\;;\quad
    \frac{Y}{2\tilde E}{q}=Y_{0}\left(\frac{c}{2}-\frac{\zeta}{2} Y_{0}[Y_{s}-Y_{0}]\right)
  \end{align}
    Putting our equations together, we have a dynamics for the two  coupled nonlinear equations of production and capital which is different from \cref{eq:21}: 
\begin{align}
  \label{eq:1700}
  \frac{d\hat Y[K]}{dt} & = Y(t)\left[\frac{\kappa-c}{2}-g_{1}+g_{2}Y_{0} \right] + Y(t)^{2}\left[\frac{\zeta Y_{s}}{2}-g_{2}\right]  
                       -Y(t)^{3}\frac{\zeta}{2} 
                          \nonumber \\ & \quad         - \frac{\kappa^{2}}{2s} \left[K(t)-K_{f}\right] + Y_{0}\left(g_{1}+\frac{c}{2}-\frac{\zeta}{2} Y_{0}[Y_{s}-Y_{0}]\right)
                         \nonumber \\ 
\frac{dK(t)}{dt} &= \epsilon_{K} s Y(t)] - \epsilon_{K} \kappa \left[K(t)-K_{f}\right]            
\end{align}
Precisely, 
the difference between these two reductions is that the quasi-stationary approximation of $E$
results in a coupling between $E$ and $Y$, \cref{eq:19}, which gives \cref{eq:21}, whereas \cref{eq:1700} is based on a coupling between $E$ and $K$.
The latter became possible only because we used the production function as an \emph{additional information} about the relation between $E$ and $K$, coupled to $Y$. 
This dynamics becomes very promising, because we can replicate the two previous scenarios: (a) damped oscillations with $Y \to Y_{0}$ and (b) sustained oscillations.
Additionally, it allows for a new stationary scenario: (c) convergence to a high production $Y\gg Y_{0}$, dependent on the choice of parameters.
This will be analyzed in the following.

To make the nonlinear equation more readable we introduce the following abbreviations for the coefficients:
\begin{align}
  \label{eq:66}
  \mathrm{CL}&=\frac{\kappa-c}{2}-g_{1}+g_{2}Y_{0}\;; \quad 
      \mathrm{CS}=\frac{\zeta Y_{s}}{2}-g_{2} \nonumber \\
          \mathrm{CQ}&=\frac{\zeta}{2} \;;\quad
          \mathrm{CC}=Y_{0}\left(g_{1}+\frac{c}{2}-\frac{\zeta}{2} Y_{0}[Y_{s}-Y_{0}]\right)
\end{align}
The cubic equation reads then in compact form:
\begin{align}
  \label{eq:67}
  \dot{Y}=\mathrm{CL}\ Y+ \mathrm{CS}\ Y^{2} -\mathrm{CQ}\ Y^{3} + \mathrm{CC} - \frac{\kappa^{2}}{2s} \tilde K(t)
\end{align}

\subsection{Bifurcation diagrams}
\label{sec:bifurcation-diagram}

\begin{figure}[htbp]
  \includegraphics[width=0.3\textwidth]{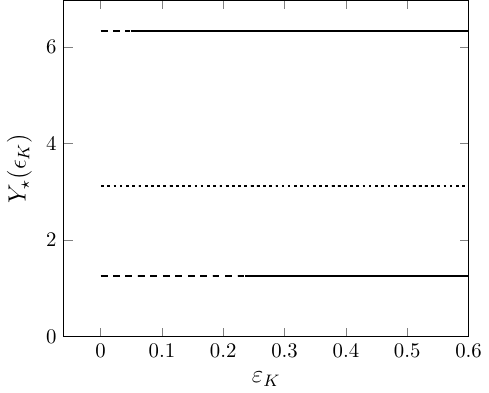}(a)\hspace*{0.0cm}    \includegraphics[width=0.3\textwidth]{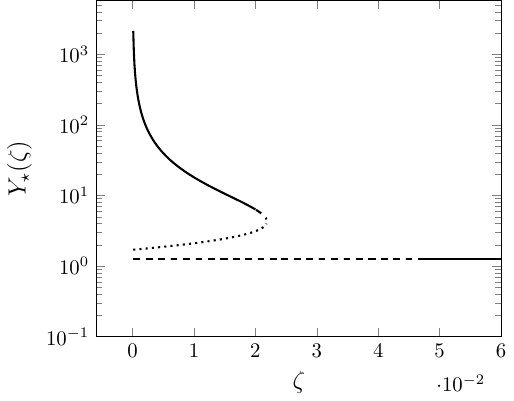}(b)\hspace*{0.0cm}
  \includegraphics[width=0.3\textwidth]{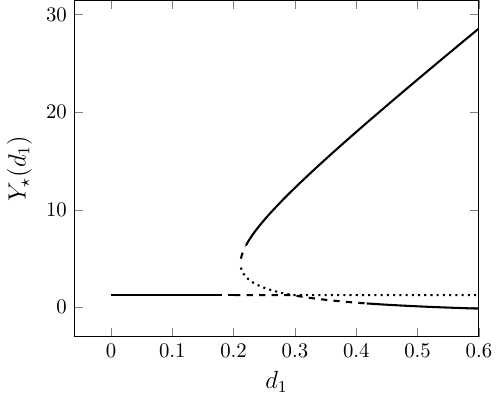}(c)

  \caption{Bifurcation diagrams of $Y_{\star}$ for different control parameters: (a) $\epsilon_{K}$, (b) $\zeta=-d_{2}$, (c) $d_{1}$. Dashed lines indicate unstable branches,  solid lines stable branches, and dotted lines saddle points. 
Parameters: 
$q_{0}$=1, $g_{1}$={0.29}, $g_{2}$=0.003, $\kappa$={0.36}, $s$={0.8}, $c$={0.06}, $d_{1}$={0.22}, $d_{2}$={0.02}, $Y_{0}$={1.25}, $K_{f}$=0, $E_{f}$=0,  $\epsilon_{K}$=0.06, }
   \label{fig:yk-bifurc}
\end{figure}

The two-dimensional dynamics defined in \cref{eq:1700} allows to calculate a bifurcation diagram from $\dot Y=0$ and $\dot K=0$.
Because of the cubic term in the production dynamics, we can expect three stationary solutions, denoted by $Y_{\star}$, $K_{\star}$.
We need to know how these solutions change if we vary control parameters of the dynamics.
We have chosen $\epsilon_{K}$ because the consideration of different time scales for the dynamics of production, energy and capital is a main feature of our model.
After eliminating $E$ from the equations, $\epsilon_{E}$ plays no role.
However, energy still implicitly impacts the dynamics of production via the parameters $d_{1}$ and $d_{2}=-\zeta$ which define $Y_{s}=-d_{1}/d_{2}$, \cref{eq:65}.
The influence of energy as a depleting resource on production is a main contribution of our paper, therefore we have chosen $d_{1}$ and $\zeta$ as additional control parameters.
\cref{fig:yk-bifurc} shows the bifurcation diagrams for the three different control parameters.

For our analysis we have chosen a set of parameters that yields three \emph{fixed points}.
For all three control parameters one of these fixed points is at $Y_{0}$ (equal to 1.25).
Additionally, we find different regimes where either three fixed points exists or only one.
The bifurcation diagram $Y_{\star}(\epsilon_{k})$ in \cref{fig:yk-bifurc}(a) clearly shows this.
It is the simplest because $\epsilon_{K}$ only affects the stability and not the values of the fixed points.
For small $\epsilon_{K}$ (below 0.048), the lower and upper fixed points are unstable and the central one is a saddle point.
In this case,
all trajectories converge to a limit cycle, similar to the one show in \cref{fig:yk-kt-0100-01679}(c) for the three-dimensional system.

Increasing the value of $\epsilon_{K}$ above 0.048, we observe  for the upper fixed point a transition from unstable to stable.
That means the former limit cycle now coexists with the stable fixed point (for $\epsilon_K<0.071$).
Trajectories originating close to the fixed point will converge to the fixed point, while other trajectories will converge to the limit cycle.
Increasing $\epsilon_{K}$ further makes the limit cycle disappearing and all trajectories converging to the upper fixed point.
For large values of $\epsilon_{K}$ (above 0.235) the lower fixed point (at $Y_0$) also undergoes transition from unstable to stable.
Depending on the initial conditions, now trajectories will converge either to the lower or upper fixed point.
The middle fixed point remains always a saddle. This means that trajectories can cross it.

The two bifurcation diagrams for the energy related control parameters $d_{1}$ and $\zeta$ are similar in that they show an inverted-$s$ curve for $Y_{\star}$.
To better understand this, we can treat $K$ as a parameter.
Doing so, $Y_{\star}$ shows three branches: the upper and lower ones are stable while the middle one unstable.
This means that all trajectories will be attracted towards these two stable branches.
We verify that both control parameters have a strong impact on the position of the \emph{upper} branch that grows with $d_{1}$ and $1/\zeta$. 

These three branches yield fixed points only when, in addition to $\dot Y=0$, also $\dot K=0$.
That means $K$ is in fact not a parameter, but a linear curve $K_{\star}=(s/\kappa)Y_{\star}$, \cref{eq:5a}, that can intersect with $Y_{\star}$ in up to three points, depending on the parameters.
There are two \emph{stable} points possible if $K_{\star}(Y_{\star})$ crosses the upper or lower branches of $Y_{\star}$, and a \emph{saddle point} where $K_{\star}(Y_{\star})$  crosses the \emph{unstable} branch of $Y_{\star}$ only once.

We take a closer look at \cref{fig:yk-bifurc}(b), starting on the right and  decreasing $\zeta$. 
For large values of $\zeta$, only one stable fixed point at $Y_{0}$ exists and all trajectories will converge to it.
Decreasing $\zeta$ below 0.047, this fixed point changes from stable to unstable and the limit cycle appears.
This transition resembles a Van-der-Pol oscillator.
At $\zeta=0.218$ a second fixed point appears which subsequently bifurcates into an upper unstable point and a middle saddle point.
Further decreasing $\zeta$ to values below 0.02, the upper fixed point undergoes a transition from stable to unstable.
The limit cycle briefly coexists with the stable fixed point, i.e., trajectories reach the limit cycle or converge to the fixed point depending on initial conditions.
Slightly decreasing $\zeta$, the limit cycle disappears and all trajectories converge to the upper fixed point whose value becomes larger with smaller $\zeta$.
Interestingly, because $Y_{0}$ remains unstable, trajectories originating below the middle fixed point first need to go below $Y_{0}$ before they can overcome the unstable point and converge to the stable one. 

The bifurcation diagram for $d_{1}$, \cref{fig:yk-bifurc}(c) has a similar interpretation.
For low values of $d_{1}$ only a stable fixed point at $Y_{0}$ exists.
Increasing $d_{1}$, we observe for this fixed point a transition from stable to unstable when the limit cycle appears, similar to a Van-der-Pol oscillator. 
For larger values of $d_{1}$ a second fixed point appears which then bifurcates into an unstable upper fixed point and a middle saddle point.
When the upper unstable fixed point undergoes a transition to a stable point, the limit cycle first coexists with the stable point.
When the limit cycle disappears, all trajectories converge to the upper fixed point.
For even larger values of $d_{1}$ the middle fixed point joins the lower point at $Y_{0}$ and then bifurcates again, yielding a saddle point at $Y_{0}$ and an unstable fixed point lower than $Y_{0}$.

For $d_{1}>0.416$ the lower fixed point changes to a stable fixed point.
That means, depending on initial conditions trajectories can converge either to a high value $Y\gg Y_{0}$ or a very low value $Y\ll Y_{0}$.
This is the regime we were interested to find: A high \emph{and} stable production.
The fact that it coexists with a regime of low and stable production illustrates the risk for the economic system.
For the same parameters the initial conditions impact whether the system ends up in a fortunate or an unfortunate regime. 

As the discussion shows, the relations between the two non-trivial manifolds resulting from $\dot Y=0$ and $\dot K=0$ generate a complex dynamics for our production model.
A particularly interesting feature is the \emph{coexistence} of stable fixed points and cycles for certain parameter ranges. 
Hence, an economic system would be stable while the macroscopic dynamics is completely different.
This coexistence of different stable solutions was already explored in other macroeconomic dynamics.
For example, for the Kaldor model a coexistence of stable and oscillatory behavior was obtained from an interplay between noise and periodic forcing \citep{Grasman-Wentzel-1994-coexistence}.
In a series of publications \citep{Dieci-Gallegati-2011-multiple,Bischi-Dieci-ea-2001-multiplekaldor,Dieci-Bischi-Gardini-2001-from} Dieci \emph{et al.} analyzed bifurcation processes that lead to the coexistence of multiple attractors, including stable equilibria and limit cycles.
They demonstrated how small changes in parameters or initial conditions can result in drastically different long-term outcomes. 
These findings have been further extended to time-discrete systems \citep{Agliari-Dieci-2006-coexistenceattractors}.

\subsection{Kaldor's model of business cycles }
\label{sec:kald-model-busin}

\begin{figure}[htbp]
  \begin{center}
    \includegraphics[width=0.3\textwidth]{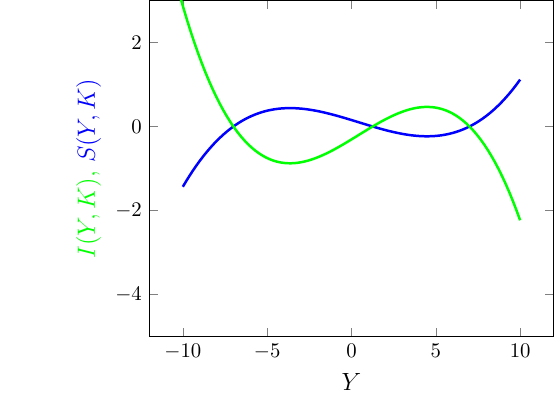}(a)\hspace*{0.0cm}
    \includegraphics[width=0.3\textwidth]{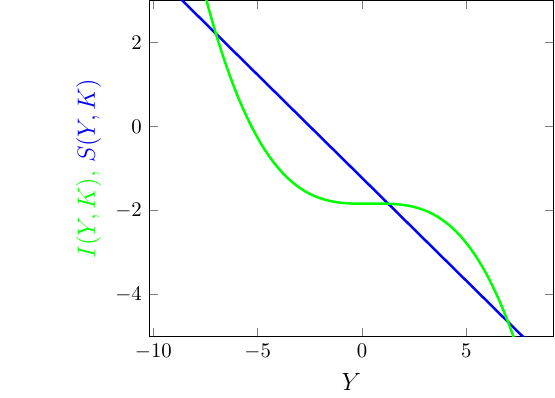}(b)\\
    \includegraphics[width=0.3\textwidth]{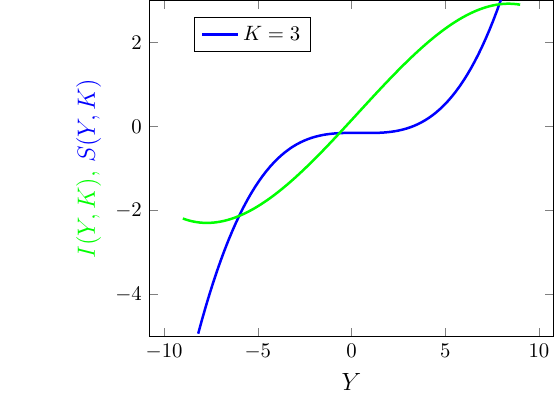}(c)\hspace*{0.0cm}
  \includegraphics[width=0.3\textwidth]{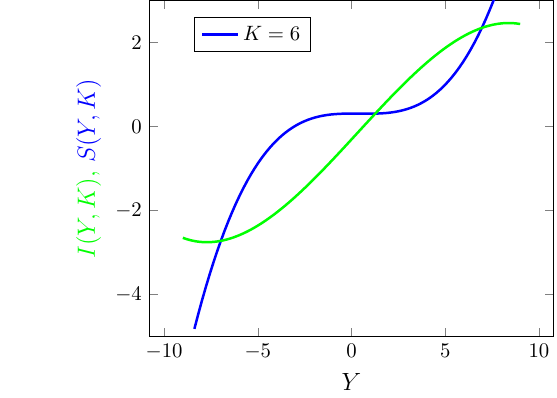}(d)\hspace*{0.0cm}
    \includegraphics[width=0.3\textwidth]{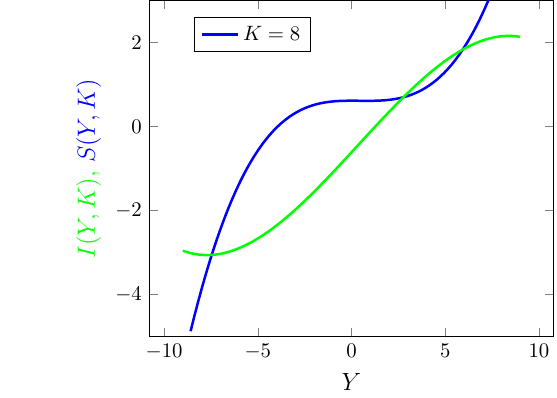}(e)
  \end{center}
  \caption{Investment $I(Y,K)$ and savings $S(Y,K)$ for different splits. (a) \cref{eq:4}, (b) \cref{eq:4a}, (c)-(e) \cref{eq:4b}. 
Parameters: 
$g_{1}$={0.01}, $g_{2}$=0.1, $\kappa$={0.7}, $s$={0.8}, $c$={0.3}, $d_{1}$={0.225}, $d_{2}$={0.02}, $Y_{0}$={3}, $K$=6.   }
  \label{fig:IS}
\end{figure}

In his famous work, \citet{Kaldor-1940-modeltrade} explains the emergence of business cycles from a mismatch between two different economic processes, investments $I(Y,K)$ and savings $S(Y,K)$, where the former is controlled by the capitalists and the latter by the workers.
He proposed the dynamics of production as follow: 
\begin{align}
  \label{eq:48}
  \frac{d Y(t)}{dt}=I(Y,K)-S(Y,K)
\end{align}
The two functions are not specified, instead they are described by a number of qualitative arguments to justify their nonlinear dependence on capital $K$ and production $Y$.
We will come back on these arguments after presenting candidates for the two nonlinear functions $I(Y,K)$ and $S(Y,K)$. 
To derive formal expressions, we need to split the nonlinear dynamics of \cref{eq:67} into two parts for $I(Y,K)$ and $S(Y,K)$ such that the sum is preserved.
This problem is not trivial because of underspecification.
We give only three characteristic examples with
the resulting functions plotted in \cref{fig:IS}.

A symmetric split would yield (see \cref{fig:IS}a):
\begin{align}
  \label{eq:4}
  I(Y,K)&=0.5\ {\mathrm{CL}}\ Y + 0.5\ {\mathrm{CS}}\ Y^{2} - 0.5\ {\mathrm{CQ}} \ Y^{3} +0.5\ {\mathrm{CC}}-0.5\ \frac{\kappa^{2}}{2s} \tilde K
\nonumber \\
                        S(Y,K)&=-I(Y,K)\;; \ \text{i.e.} \dot{Y}=2\ I(Y,K) 
\end{align}
A linear assumption for $S(Y,K)$ leads to (see \cref{fig:IS}b):
\begin{align}
  \label{eq:4a}
I(Y,K)&=\mathrm{CS}\ Y^{2}- \mathrm{CQ}\ Y^{3}- \frac{\kappa^{2}}{2s}\tilde K \nonumber \\
  S(Y,K)&=-\mathrm{CL}\ Y - \mathrm{\mathrm{CC}} 
\end{align}
An uneven split of the nonlinear terms between $I(Y,K)$ and $S(Y,K)$ reads instead (see \cref{fig:IS}c,d,e):
\begin{align}
  \label{eq:4b}
I(Y,K)&= \mathrm{CL}\ Y + 0.2\ \mathrm{CS}\ Y^{2} - 0.25\ \mathrm{CQ}\ Y^{3} +0.5\ \mathrm{CC} -0.5\ \frac{\kappa^{2}}{2s}\tilde K \nonumber \\
  I(Y,K)&= -0.8\ \mathrm{CS}\ Y^{2} + 0.75\ \mathrm{CQ}\ Y^{3} + 0.5\ \mathrm{CC} + 0.5\ \frac{\kappa^{2}}{2s}\tilde K
\end{align}
All three variants lead to the same dynamics for $Y(t)$, \cref{eq:67}.
In particular all of them have the same stationary solutions.
Thus, the only way to make a meaningful choice is to resort on \emph{economic arguments}, keeping in mind that \emph{investments} $I(Y,K)$ and \emph{savings} $S(Y,K)$ have an economic meaning.
Here, we refer to the work of Kaldor \citep{Kaldor-1940-modeltrade}, but do not repeat the details of the argumentation.

First, Kaldor argues that both functions have to \emph{monotonously increase} with production $Y$.
This rejects \cref{eq:4} (\cref{fig:IS}a) because these  functions are \emph{non-monotonous} in $Y$.
It also rejects \cref{eq:4a} (\cref{fig:IS}b) because these functions are monotonously \emph{decreasing} in $Y$.
The variant of \cref{eq:4b} (\cref{fig:IS}c,d,c) however fulfills this requirement.

Second, Kaldor distinguishes the behavior of the two functions in case of \emph{high} and \emph{low} capital.
If capital is \emph{low}, investments should increase with $K$ over time to make use of many good investment opportunities.
Savings, however, should decrease with $K$ over time because of rising prices.
On the other hand, if capital is \emph{high}, investments should decrease, while savings should increase.
This implies that both curves move against each other as indicated in \cref{fig:IS}(c,d).
If $I(Y,K)$ moves up, $S(Y,K)$ moves down (for low levels of $K$) and the other way round (for high levels of $K$).
This requirement is met by the functions of \cref{eq:4} and \cref{eq:4b}, but not by \cref{eq:4a}, because here $S(Y,K)$ does \emph{not} depend on $K$ and therefore does not move.
Nevertheless, \cref{eq:4a} would generate oscillations like the other examples.

We conclude that our proposal for the two nonlinear functions given in \cref{eq:4b} fulfills Kaldor's requirements.
It should be noted, however, that the concrete shape of our nonlinear functions depend on the chosen parameters, thus statements about the slope and the monotonous increase are restricted to this choice.

Kaldor \citep{Kaldor-1940-modeltrade} argues that two \emph{linear} functions could still monotonously increase with $Y$ and correctly depend on $K$, but they would only allow for one stationary solution.
This solution would be \emph{unstable} if the slope of $I(Y,K)$ is larger than the slope of $S(Y,K)$.
Hence, the economy would either only grow or only shrink.
On the other hand, if the  slope of $I(Y,K)$ is smaller than the slope of $S(Y,K)$, the stationary solution would be \emph{stable} and the economy would remain around this stable state.
This, however, contradicts the broad observations of \emph{business cycles}, which Kaldor wanted to explain with his investigations.

\subsection{Relation to the Van-der-Pol oscillator}
\label{sec:relation-van-der}

Our model generates business cycles and provides functional expressions for the dynamics.
Reproducing the asymmetric duration of the phases, however, is not the merit of our efforts, it shall be attributed to the underlying general dynamics of a Van-der-Pol oscillator. 
This oscillator is a paragon of a non-linear system with \emph{nonlinear} friction.
To compensate this friction, oscillations require the input of energy.
That means, we have a \emph{dissipative} system and the oscillations can be classified as \emph{limit cycles}.
For a critical energy supply, we obtain a Hopf bifurcation.

The Van-der-Pol oscillator can be formalized with one second-order or with two coupled first-order differential equations.
For our comparison the latter is more suitable. 
\begin{align}
  \label{eq:420}
  \frac{dy(t)}{dt}&=\omega\left[y(t)-\frac{1}{3}y^{3}(t)-x(t)\right] \nonumber \\
  \frac{dx(t)}{dt}&=\frac{1}{\omega}\,y(t)
\end{align}
Different from our two-dimensional dynamics, \cref{eq:1700}, the Van-der-Pol oscillator has only one control parameter $\omega$ that appears in both equations.
To obtain oscillations $\omega>0$ is required.
Only for large $\omega$, these oscillations remind of business cycles with short and long phases.
To map the equations of the Van-der-Pol oscillator back to the Kaldor model, \cref{eq:48}, we first have to drop the depreciation term $\kappa K$ in the dynamics of capital, \cref{eq:25}, i.e., $\dot{K}=I$.
If $\dot{Y}=\alpha(I-S)$ and $\dot{K}=I$ is then solved for $I(Y)$ and $K(Y)$, using \cref{eq:420}, one finds \citep{Chian-2007-complexsystems} 
\begin{align}
  \label{eq:1}
  I(Y)&= \frac{1}{\omega}Y \nonumber\\
  S(Y)&= \frac{1-\omega}{\omega}Y+\frac{Y^{3}}{3}+K
\end{align}
where $\omega$ is a factor of $\alpha$. 
Note that $S(Y,K)$ is proportional to $K$, similar to our \cref{eq:4b}, while $I(Y,K)$ does not depend on $K$, against Kaldor's requirements.
While convenient for nonlinear dynamics, 
the fact that we only have one control parameter $\omega$ is considered a drawback 
for economic applications because we cannot distinguish the different processes underlying economic dynamics.
\cref{eq:4b} instead provides more flexibility.

\section{Discussion}
\label{sec:discussion}

There is an ongoing debate about the origin of business cycles.
Are they exogenous, triggered by external shocks such as wars, natural catastrophes, political collapse?
Or, are they endogenous, that means, resulting from the internal dynamics of a country's economy, changes in consumption, inflation?
The answer is probably that both endogenous and exogenous causes play a role.
No surprise if \emph{big} disruptions, like a major earthquake destroying production sites, induce a recession, even a depression of economic activities.  
From a modeling perspective it is more interesting how \emph{small} disruptions, e.g., the failure of single elements, can be amplified such that a whole system collapses \citep{Lorenz2009}.
This requires to understand the internal feedback mechanisms that generate the system dynamics \emph{endogenously}.
In this paper, we provide a minimal model that allows to study such endogenous effects on  economic growth, in particular the appearance of business cycles, in a systematic manner. 
    Unlike previous models, which often relied on external shocks or specific nonlinear functions, our model demonstrates how these complex dynamics can arise endogenously from the interactions between production and resource constraints.

The starting point of our investigations was to propose a production function that, in addition to capital, depends on a generalized ``energy'': a resource that is depleted during production.
Increasing production thus means decreasing energy.
Our production dynamics results from this production function, together with assumptions about the dynamics of capital and energy.
Here, we have chosen an ansatz that considers different time scales and a nonlinear dependence of production on energy.
This way, we have derived a nonlinear dynamics of production that is able to generate  business cycles endogenously, i.e., without assuming external shocks or time lags.   
Under the constraint of a specific coupling between capital and energy, we find additionally fixed points of the dynamics where the production is considerably higher than a baseline.
While both cycles and fixed points are stable, our model is sensitive to parameter changes and initial conditions.
This should be seen as a an advantage because in complex systems already small deviations can lead to instabilities, and economic systems are no exception. 

The fact that energy is consumed during production drives the economy out of a thermodynamic equilibrium, which is reflected in our model.
The take-up of energy, its transformation and dissipation are features of ``active'' matter
  \cite{Schweitzer-2019}. 
Energy enables systems to evolve and to self-organize \citep{Feistel-Ebeling-2011-physicsself}.
But in social and economic systems this does not imply  a desired outcome, as witnessed by business cycles in economic activities. 

As a modeling consequence of such out-of-equilibrium situation, we have to distinguish between driving and driven variables.
In our case, energy is the driving variable.
Its take-up and transformation increases production as the driven variable.
This bears similarities to active motion, the directed movement of biological entities such as cells or animals~\citep{Ebeling2003}. 
In both cases, energy is consumed, that means \emph{decreased} to increase production or speed.
This makes our model different from related models of business cycles, which do not reflect the consumption of energy.
The constrained resource sets limits to the increase of production, resulting in a saturated growth dynamics.
Our model considers, in addition to the driven dynamics, also the \emph{eigendynamics} of the system.
That means, it captures the production dynamics in the \emph{absence} of additional input, a feature not addressed in simpler macroeconomic models.
It is in fact the interplay between eigendynamics and driven dynamics that generates non-trivial stationary solutions, such as high stable production or limit cycles.
The coexistence of stable fixed points and limit cycles, already observed in other models of business cycles, is particularly interesting.
It allows to discuss intervention mechanisms that can not only stabilize economic dynamics, but also drive it to preferred states \citep{schweitzer-2020-design}, offering a new perspective on business cycle dynamics.

\small \setlength{\bibsep}{1pt}

\end{document}